\pdfoutput=1
\documentclass{bmcart}
\usepackage[utf8]{inputenc}
\usepackage{graphicx}
\usepackage{amsmath}
\UseRawInputEncoding

\startlocaldefs
\endlocaldefs

\begin{document}

\begin{frontmatter}

\begin{fmbox}
\dochead{Research}
 
\title{Topology-dependence of propagation mechanisms in the production network}

\author[
   addressref={aff1, aff2, aff3},  
   email={e.molnar@lancaster.ac.uk} 
]{\inits{EM}\fnm{Eszter} \snm{Moln\'{a}r}}
\author[
   addressref={aff1,aff3, aff4},
    corref={aff1}, 
   email={d.csala@lancaster.ac.uk}
]{\inits{DCS}\fnm{D\'{e}nes} \snm{Csala}}

\address[id=aff1]{
  \orgname{Department of Engineering, Lancaster University}, 
 \city{Lancaster},                            
  \cny{UK}                                    
}
\address[id=aff2]{%
  \orgname{Material Social Futures Doctoral Training Centre, Lancaster University},
  \city{Lancaster},
  \cny{UK}
}
\address[id=aff3]{%
  \orgname{Faculty of Economics and Business Administration, Babes-Bolyai University},
  \city{Cluj-Napoca},
  \cny{Romania}
}
\address[id=aff4]{%
  \orgname{Economics Observatory, School of Economics, University of Bristol},
  \city{Bristol},
  \cny{UK}
}

\begin{artnotes}
\end{artnotes}

\end{fmbox}

\begin{abstractbox}

\begin{abstract} 
The topology of production networks determines the propagation mechanisms of local shocks and thus the co-movement of industries. As a result, we need a more precisely defined production network to model economic growth accurately. 
In this study, we analyse Leontief's input-output model from a network theory perspective, aiming to construct a production network in such a way that it allows the most accurate modelling of the propagation mechanisms of changes that generate industry growth. We do this by revisiting a prevalent threshold in the literature that determines industry-industry interdependence. Our hypothesis is that changing the threshold changes the topological structure of the network and the core industries to a large extent. This is significant, because if the production network topology is not precisely defined, the resulting internal propagation mechanisms will be distorted, and thus industry growth modelling will not be accurate. 
We prove our hypothesis by examining the network topology, and centrality metrics under different thresholds on a network derived from the US input-output accounts data for 2007 and 2012.  

\end{abstract}


\end{abstractbox}

\end{frontmatter}

\section*{1 Introduction}
No industry can exist and develop individually. All sectors are interdependent through the exchange of products and services \cite{xu_inputoutput_2019}. Given the increasing complexity of our society, we cannot examine a problem without the help of interdisciplinary approaches. Inter-industry relations are profoundly interdisciplinary; therefore, we must use techniques that view industries from a systematic perspective and understand the complexity of relations. In this study, we use network and data science tools to explore industry inter-relatedness through the help of input-output analysis.

From economics perspective an industry is a branch of an economy that produces a closely-related raw materials, goods, or services, while from business point of view it is a group of companies that are related based on their primary business activities. There are dozens of industry classifications, and these classifications are typically grouped into larger categories called sectors.

Leontief's \cite{leontief_environmental_1970, dietzenbacher_wassily_2008} economic input-output models represent, in mathematical form, the monetary transactions between industry sectors. They specify what goods and services (output) are consumed by other industries (input).

The network science approach to input-output models is not a novel concept. There are mainly two basic approaches currently being adopted in this research area. One is the analysis from a supply chain perspective, using company-level data \cite{wu_centrality_2015, brintrup_supply_2018, perera_network_2017}, and the other is the industry perspective, using country or global-level input-output accounts.

In the industry perspective approach, a considerable amount of literature has been published using the World Input-Output Database \cite{xu_inputoutput_2019, soyyigit_global_2018, grazzini_empirical_2015, soyyigit_input-output_nodate}, covering 40 countries in the 2013 release and 43 countries in the 2016 release, all with 35 (2013) and 56 (2016) sectors. \cite{timmer_illustrated_2015} Although this data source can cover most countries, it only contains information on very aggregated sectors. In recent years, researchers have also investigated various approaches to the input-output transaction data of the US economy as systematised by the Bureau of Economic Analysis (BEA) \cite{input_output_bea}. Most of the studies focused on the sector and summary level of the input-output accounts containing 21 (sector-level) and 71 (summary-level) aggregated industries \cite{foerster_changing_2017, duan_modelling_2012}.

Sectoral inter-dependencies are pivotal in connecting microeconomic shocks with business fluctuations and cycles, especially during the supply-chain fluctuations of today. Researchers claim that production networks derived from input-output models provide a valuable account in opening the black box of co-movement and propagation mechanisms that shape aggregate outcomes. Carvalho \cite{carvalho_micro_2014} points out that production network research can advise scholars on the origins of aggregate fluctuations and policymakers on how to be ready and recuperate from disadvantageous shocks that disturb production chains \cite{keith_2017, dykes_dynamics_2017}.

Horvath \cite{horvath_cyclicality_1998, horvath_sectoral_2000} and Acemoglu et al. \cite{acemoglu_network_2012} argue that the structure of the production network plays a crucial role in determining the aggregate behaviour of the system. Network structure is the set of nodes and edges of a network. Nodes representing industries in the production network, carry certain quantitative properties, which are represented as weights. The edges representing the monetary transactions between industries also carry weights, but in the case of a production network they are also directed - and the weights vary significantly in terms of directions.

When the production network is outstandingly asymmetric, for example, when few sectors are in a dominant role as suppliers, idiosyncratic shocks lead to aggregate fluctuations. If the production organisation is dominated by a small number of hubs supplying inputs to many different sectors, disturbances at these crucial nodes will affect the global production system, determining losses in production and welfare \cite{acemoglu_network_2012}. 
Bigio and La'O \cite{bigio_financial_2016} also demonstrated that overall network topology defines the strength of each channel. They compared two production networks: a horizontal and a vertical economy, and showed that the network centrality of sectors matters for how they affect aggregate output. Carvalho \cite{carvalho_micro_2014} also advocates for the significance of network topology by comparing the amplification of micro-level volatility and the network multiplier a horizontal economy with no input trade, a vertical economy with a source and a sink, and a star/hub-and-spoke economy with a central node/s.
 
The US Bureau of Economic Analysis publishes the Industry Economic Accounts generally at three levels of detail: sector (21 industry groups), summary (71 industry groups), and detail (405 industry groups).
For example at the sector-level Durable goods is included between the 21 industry groups. This sector at the summary-level contains Primary metals, Machinery, Computer and electronic products and 8 other sub-sectors. While broken down even further to detail-level, the Primary metals include 10 industries (Iron and steel mills and ferroalloy manufacturing, Ferrous metal foundries, etc.), the Machinery includes 28 industries (Farm machinery and equipment manufacturing, Semiconductor machinery manufacturing, etc.), and Computer and electronic products include 20 industries (Electronic computer manufacturing, Telephone apparatus manufacturing, etc.).
These industry classifications are all grouped hierarchically in three levels.

Although extensive research has been carried out from an industry perspective, just a few studies exist which develop a network of at least 400 detail-level industries. This can be a key problem because the networks built on a highly aggregated level with just a few nodes and connections might not represent the industry interdependencies accurately. 

On the one hand, the topology of a detail-level industry network can differ considerably from an aggregate-level network. The first one tends to be way denser with more lower-weighted connections, thereby behaving differently. On the other hand, some essential links could be hidden in an aggregate-level network. For example, embedded in a highly-weighted connection, several detail-level links could have been hidden that might be more important than the other present aggregated ones. The whole map of industrial interdependence could change when analysing these separately. The detail-level input-output account data might allow us to discover a more representative picture not just with more separable sectors, but with way more supporting connections between industries, in number and validity. This increased granularity of understanding allows for a better identification of key industries undergoing change and more efficient tracking and understanding of innovation.

Most scholars, like Carvalho, use a threshold value and consider only a percentage of the monetary transactions to be present in the production network. As the monetary transactions will become the edges between industry-industry pairs when constructing the production network, they do this mainly to make the data more manageable. Initially, in the input-output models, there are almost $i^2$ supporting transactions, with \emph{i} being the number of industries. The production networks obtained from such detail-level input-output models are exceptionally dense and challenging to analyse, without offering much insight. The scholars' solution is to impose a threshold and exclude a percentage of the smaller monetary transactions - thus "severing" these links in the production network. In our study we refer to this cut-off point as threshold $\zeta$.

Collectively, these studies outline a critical need to examine how the threshold $\zeta$ influences the topology of the production network, and thus the propagation mechanisms and aggregate fluctuations. While we might agree that production networks can provide a bridge between micro and macro \cite{carvalho_micro_2014}, very little is currently known about how much the threshold value $\zeta$ chosen by scholars distorts this bridge.

In this study, we discover the production network's considerable structural volatility to the threshold value $\zeta$ by analysing a detail-level input-output model with a high number of sub-sectors, to understand the most accurate structure of the network of industries. 
The Data and methodology section presents the chosen dataset, the production network model, and the topological and node-level analysis.

\section*{2 Data and methodology}
\subsection*{Data source}
We used the detail level United States Input-Output Accounts Data \cite{input_output_bea} to conduct this research. 
The Bureau of Economic Analysis publishes the Industry Economic Accounts generally at three levels of detail: sector (21 industry groups), summary (71 industry groups), and detail (405 industry groups). Estimates at the detail level are produced roughly every five years, with the last two releases being from 2007 and 2012. While there are Historical Benchmark Input-Output Tables from 1947 until 2002, these reflect industry definitions that vary across years, and BEA advises that they should not be used as a time series.
BEA uses its Industry Codes for the three levels of detail, but it also defines how these relate to the 2012 North American Industry Classification System (NAICS) code structure \cite{naics}.

We used the last two releases, the 2007 and 2012 Total Requirements table - Industry by Industry, including 405 industries and inter-industry purchases. Additional explanations regarding the Total Requirements derivation can be found on BEA's website \cite{total_requirements}.

\subsection*{Production network generation from data}
The BEA Total Requirements table had a network representation in which an element $w_{ij}$ represented the nominal amount of goods \emph{i} used as input by sector \emph{j}, with \emph{i, j} = 1, ..., \emph{N}, where \emph{N} was the number of sectors.

For a two-industry example of the Total Requirements table, see Table~\ref{table:input-output}. The network built from this database was a directed weighted graph. The vertices were the industries (\emph{i} and \emph{j}), the directed connections were the monetary transactions, the nominal flow of goods between sectors. The weight of each link was the economic value, representing how much an industry supports the development of the other ($w_{ij}$ and $w_{ji}$). For example industry \emph{i} requires $w_{ij}$ dollar input from industry \emph{j} to produce one dollar output to final users (typically $w_{ij}$$\neq$$w_{ji}$). 

The input-output models also indicate if output of an industry is required as input to the same industry ($w_{ii}$). For example the Oil and gas extraction industry could produce the oil and gas to power its own equipments, or the computer design and manufacturing industry can produce computers that are used to plan the next generation of computers \cite{eiolca}. We didn't use these self-sector transactions in our analysis.

\begin{table}[ht!]
\caption{Economic Input-Output Model}
      \begin{tabular}{cccc}
        \hline
           & Industry i  & Industry j\\ \hline
        Industry i & $w_{ii}$ & $w_{ji}$\\
        Industry j & $w_{ij}$ & $w_{jj}$\\ \hline
      \end{tabular}
      \label{table:input-output}
\end{table}

\subsection*{Threshold $\zeta$}
After building the production network from the Input-Output Accounts Data, we defined one hundred different edge cut-off thresholds $\zeta$ from 0.00001 to 0.5 with 0.005 equal intervals. The threshold $\zeta$ was associated with the value of the links, representing inter-industry trade in monetary expression. For example, at the end, when the threshold value $\zeta$ was 0.5, we only considered those connections that weighed at least 0.5. In other words, only those monetary transactions become edges in the resulting production network where at least half a dollar is needed from one industry to produce one dollar output in another. At this threshold $\zeta$, in the 2012 production network, only 19 industries (nodes - in the 2007 network, 17 industries) and 11 monetary transactions (edges) were present. We chose this long interval and these small steps to represent the best topology change according to the cut-off $\zeta$.

\subsection*{Algorithms for random and scale-free models}
Then, we \textit{pruned} the graph according to the considered cut-off thresholds $\zeta$. We generated a random and a scale-free graph with the same parameters as our production network at each cut-off point $\zeta$. For both, we used the NetworkX graph generator algorithms \cite{networkx}. We gave the same node and edge number for the random graph generator algorithm as it was in our original production network. For the scale-free graph, we used the algorithm implemented after Bollobás et al. \cite{bollobas_directed_2003} with the same node number and by calculating $\alpha$, $\beta$, $\gamma$ and $\delta$ parameters to fit our production network.

\subsection*{Degree distribution comparison (Kolmogorov-Smirnov test)}
We chose the Kolmogorov-Smirnov (KS) statistical test to compare our observed degree distribution to what we would expect from a random and a scale-free graph with the same parameters, because it is widely used, when comparing degree distributions for networks. \cite{deng_exponential_2011, muchnik_origins_2013, gomez_statistical_2008, gjermeni_temporal_2017} 

As our production network is a directed network, we distinguished between in-degree and out-degree: 
\begin{equation}
k_{in, i} =\sum_{j=1}^{N} a_{ij}  \hspace{1cm} k_{out, i} =\sum_{j=1}^{N} a_{ji}
\end{equation}
where $k_{in, i}$ is the incoming degree of node \emph{i}, representing  the number of incoming edges onto the node, $a_{ij}$ is 1 if there is a directed transaction between industry \emph{i} (buyer) and industry \emph{j} (seller). In other words, if industry \emph{i} requires the output of industry \emph{j}, otherwise it is 0. The outgoing degree is $k_{out, i}$, representing the number of links that point from node \emph{i} to other nodes. In this case, $a_{ji}$ is 1, if the industry \emph{i} is a supplier to industry \emph{j}. In these measures self-sector transactions, indicating if output of an industry is required as input to the same industry, are not included ($a_{ii}$).

We analysed and compared in-degree and out-degree distribution separately at each threshold $\zeta$ by observing when does the network behave like a random or scale-free graph. We used the SciPy two-sample KS statistical test based on Hodges \cite{hodges_significance_1958}. The KS statistic quantified the distance between the empirical distribution functions of the two samples (the observed degree distribution and the random/scale-free graph's degree distribution). The smaller this value was, the more likely the two samples were drawn from the same distribution.

\subsection*{Centrality measures}
We calculated two network centrality measures for every node at six cut-off points $\zeta$ from 0.0 to 0.1 with 0.02 equal intervals on the directed weighted graph. 
In this interval, monetary transactions are exponentially disregarded, and at 0.1 there are still more connections than nodes in the production network. With 6 cut-off points $\zeta$ in this area we can illustrate well the change in the central industries.
Degree centrality is defined as the total number of links of a node. The production network is weighted and directed; thereby, we computed the weighted in-degree (in-strength) and the weighted out-degree (out-strength) of each node:
\begin{equation}
k_{in, i}^{w} =\sum_{j=1}^{N} w_{ij}  \hspace{1cm} k_{out, i}^{w} =\sum_{j=1}^{N} w_{ji}
\end{equation}
where $k_{in, i}^{w}$ is the in-strength of industry \emph{i}, representing the sum of all incoming flows of goods, the nominal inputs used by the sector \emph{i}, $w_{ij}$ is the transaction value between industry \emph{i} (buyer) and industry \emph{j} (seller). In other words the required amount from industry \emph{j} to produce one dollar of output from industry \emph{i}. The outgoing degree for industry \emph{i} is $k_{out, i}^{w}$, representing the sum of the outflow of goods from node \emph{i} to other nodes. In this case, $w_{ji}$ is nominal input that industry \emph{i} supplies to industry \emph{j}. We didn't use self-sector transaction weights representing in what amount the output of an industry is required as input to the same industry ($w_{ii}$).

We chose the weighted out-degree and the weighted \textit{PageRank} out-degree centrality \cite{page_pagerank_1999, langville_survey_2004} as our key metrics. The first represented how big and indispensable an industry is in monetary terms, \textit{globally} in the entire production network, and the second represented how key an industry is in terms of \textit{location} in the network. The reason behind the PageRank algorithm is that a node is systemically important if its neighbours are important and/or the neighbours of the neighbours are important. 
We computed the PageRank on the reversed graph to get the weighted PageRank out-degree centrality. At each threshold $\zeta$, we showed the top 20 central industries and compared how they changed according to the cut-off.

Besides this, we also compared the weighted in-degree and out-degree distribution without a cut-off $\zeta$ to dig deeper into the asymmetry analysis.

\section*{3 Results}
Our results provide a \textit{first} insight into the production networks' structural volatility to the threshold value $\zeta$. We first analysed the topological features and found that the industry inter-dependence network topology is highly sensitive to the chosen threshold $\zeta$. Starting with the smaller monetary value transactions, as we disregarded the higher and higher ones, the production network's topology, from the out-degree perspective, went through a 'scale-free to random' topology shift.

\begin{figure}[ht!]
    \includegraphics[scale=0.36]{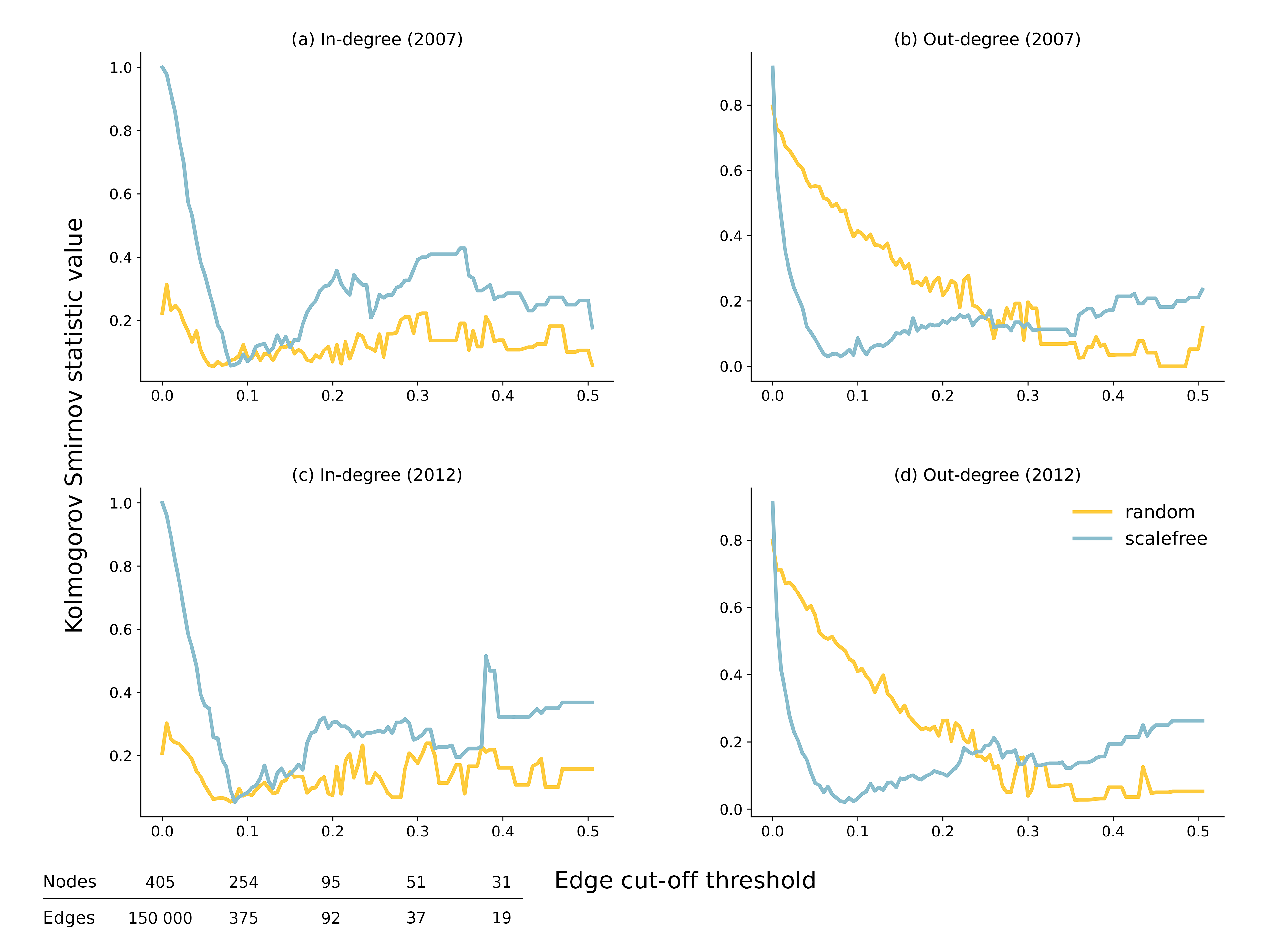}
  \caption{\csentence{Kolmogorov-Smirnov statistic value of the distributions compared}
      This figure shows 4 line charts that show the KS statistical test's value at 100 edge-cut off thresholds $\zeta$ when comparing the production networks' observed degree distribution to what we would expect from a random and scale-free network with the same parameters, according to year and degree type. The closer the KS statistic value is to 0, the more likely the two samples were drawn from the same distribution. The node and edge numbers are also included according to 5 cut-off points $\zeta$.}
  \label{fig:KStest}
\end{figure}

Figure~\ref{fig:KStest} shows the KS statistical test's value at various edge cut-off thresholds $\zeta$. We compared the production networks' observed in-degree and out-degree distribution at each threshold $\zeta$ to what we would expect from a random and scale-free network with the same parameters. The closer the KS statistic value was to 0, the more likely the two samples were drawn from the same distribution. 

With the threshold $\zeta$ increasing, the node and edge numbers declined exponentially. According to out-degrees, the 'scale-free to random' topology shift appeared around the 0.25 threshold value ($ \zeta \approx 0.25 $) when fewer than a hundred nodes and edges were present in the production network (70 nodes and 60 edges). The topology shifted completely when 15\% of all nodes and 0.05\% of all edges were included. Even until this point, it showed some changes, despite being still in the scale-free range. Most scholars accounted for about 80\% of the total value of input trade, which would leave us with 5000 edges, 3.5\% of all transactions in the network, around the 0.02 cut-off point $\zeta$ ($\zeta \approx 0.02 $). While others considered 1000 edges, 0.7\% of all transactions, at the 0.05 cut-off point $\zeta$ - and at this moment, we have already lost 30 industries out of four hundred. 
The in-degree network had a topology more similar to a random network than a scale-free network, regardless of its edge cut-off threshold $\zeta$. 

\begin{figure}[ht!]
  \includegraphics[scale=0.38]{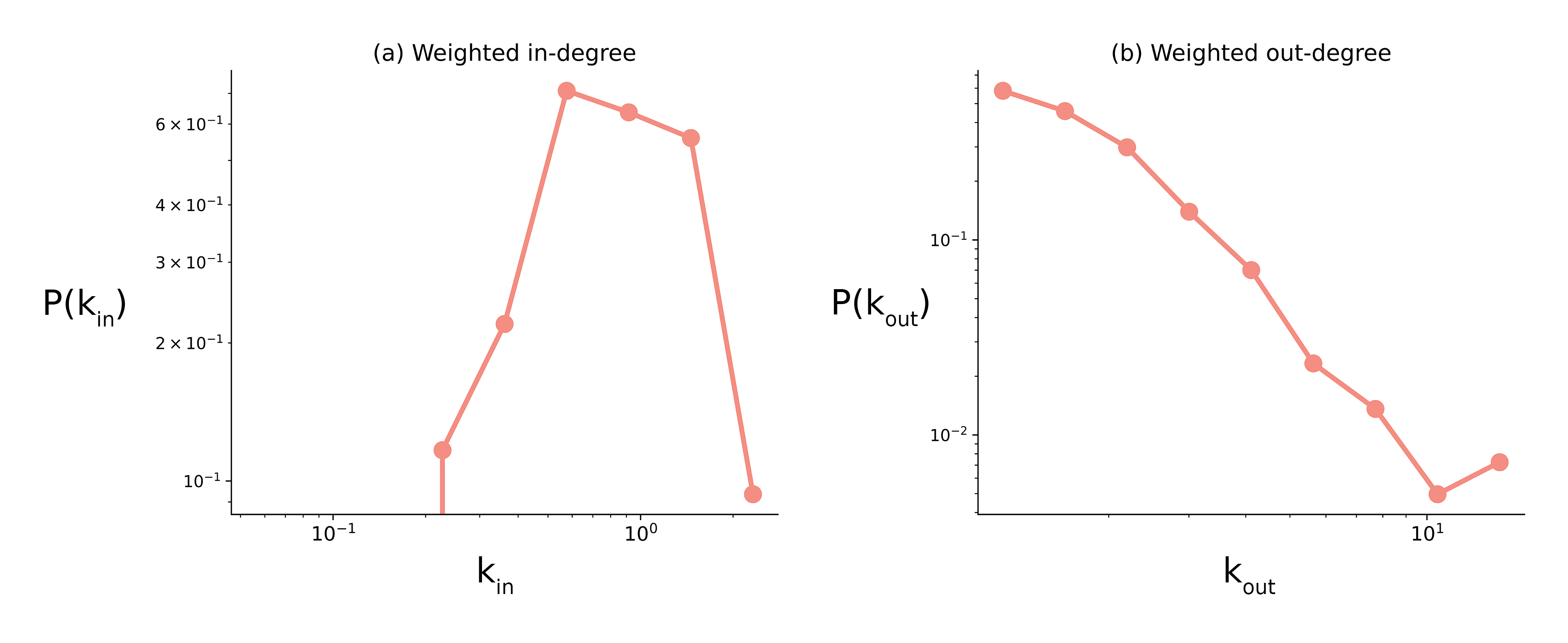}
  \caption{\csentence{Degree distribution of the production network}
      The (a) in-degree and (b) out-degree distribution of the entire production network without cut-off in log-log scale and by considering the amount of each monetary transaction (the weights of the edges).}
  \label{fig:DegreeDist}
      \end{figure}

Our results show that the in-degree and out-degree distributions of the production network behave very differently. It is more similar to a random network from the in-degree perspective, whereas, from the out-degree perspective, it leans towards a scale-free topology. To discover more about this asymmetry, we showed the weighted in-degree and out-degree distribution of the entire production network without cut-off $\zeta$ in Figure~\ref{fig:DegreeDist}. 
Some scholars already approached the question of asymmetry in directed networks \cite{luo_asymmetry_2015, wang_influences_2017}. For the case of the production network was quite transparent the underlying explanation.

The topological difference came from the presence of critical sectors that dominated economic activity, the so-called "commanding heights". Vladimir Lenin used this phrase in the early 1920s, referring to the control of key segments of a national economy. The difference between raw and processed materials was evident in the degree distribution comparison. The core industries that drive the economy formed hubs and pushed the topology to the scale-free range from the out-degree perspective. While from the in-degree perspective, most sectors needed the same amount of resources; therefore, the topology leaned towards a random network.

\begin{figure}[ht!]
\includegraphics[scale=0.61]{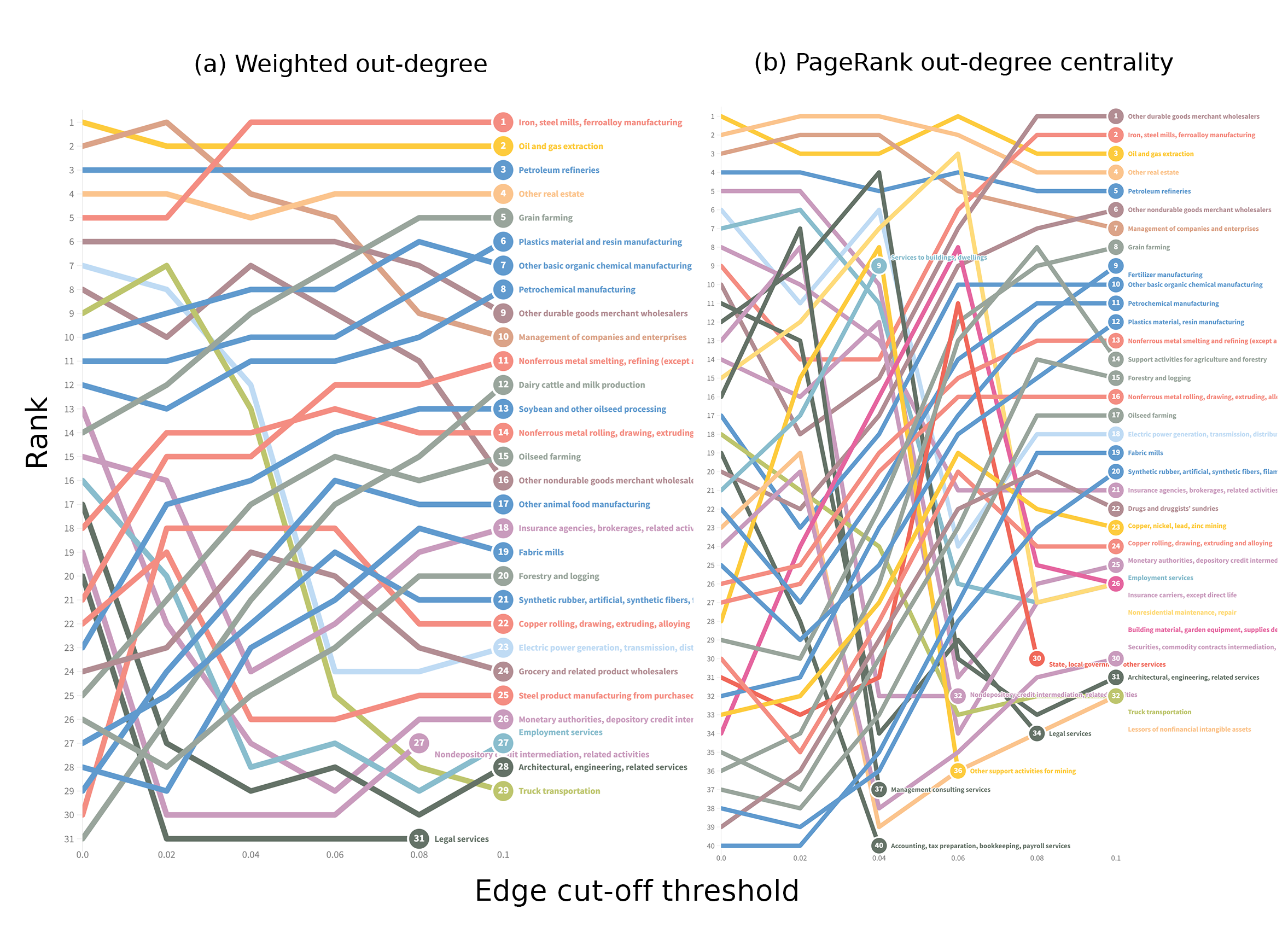}
  \caption{\csentence{Top industries according to threshold $\zeta$}
      This figure shows the node-level threshold $\zeta$ sensitivity of the production network. We reveal the top 20 sectors according to (a) weighted out-degree and (b) PageRank out-degree centrality at six cut-off points $\zeta$.}
  \label{fig:NodeLevel}
      \end{figure}

According to size and centrality measures, our results show that some core industries tended to remain on top at each threshold $\zeta$. Figure~\ref{fig:NodeLevel} reveals the node-level threshold $\zeta$ sensitivity of the production network. We calculated the top 20 sectors according to weighted out-degree and weighted PageRank out-degree centrality at six cut-off points $\zeta$. We found that the top three industries stayed the same: 1. Iron, still mills, ferroalloy manufacturing, 2. Oil and gas extraction, 3. Petroleum refineries. 

Because the threshold $\zeta$ change cut off the minor value transactions, some industries leading the top lists have dropped to the bottom of the rankings. These were usually the industries that supported almost all the other ones, but with smaller transactions, such as Truck transportation and Electric power generation, transmission, distribution. Therefore, we can conclude that the threshold $\zeta$ cuts out some significant hubs that amplify aggregate fluctuations.

\section*{4 Discussion}
In this study, we aimed to construct a production network for an economy so that it allows the most accurate modelling of the propagation mechanisms of changes that generate industry growth. 

Our hypothesis was that changing the percentage of monetary transactions that researchers include in the network (edge cut-off threshold $\zeta$) changes the structure of the network and the core industries to a large extent. This was significant because if the production networks' structure is not precisely defined, the resulting internal propagation mechanisms will be distorted, and thus industry growth modelling will not be accurate.
We proved our hypothesis by examining the network topology and centrality metrics under different thresholds $\zeta$ on a network derived from the US input-output accounts data for 2007 and 2012. 

The topological analysis's outcome enabled us to define the propagation mechanisms in the production network. By demonstrating that it is scale-free from the out-degree perspective, we showed that it has hub sectors, the so-called "commanding heights" that strengthen the internal propagation mechanisms as they are. We also discovered that the industry inter-dependence network is susceptible to the edge cut-off threshold $\zeta$. The production network's topology went through a 'scale-free – random' topology shift, as we disregarded monetary transactions increasing in number and size. This meant that the propagation mechanisms depended on the threshold value $\zeta$ chosen just as the structure.  

The node-level threshold $\zeta$ volatility analysis showed us that not just the overall topology changes but some core industries that served as hubs dropped to the bottom of the top lists as we decreased the number of transactions included. These industries mainly supported numerous other industries with small monetary transactions.

In his study, Carvalho \cite{carvalho_micro_2014} mentions that hubs shorten distances. While cutting out these transactions, we lose these hubs and increase distances.
These monetary transactions contribute directly to the production networks' connectedness and complexity, which explains the propagation (and sometimes circularity) of local shocks and disturbances. A reasonable question to ask would be, if we do not consider them all, in what proportion will our industry growth modelling be distorted?
	
For example, even small supporting transactions could make a difference. While we might cut out most of the Truck transportation transactions, a disturbance in the industry could impact all the others dependent on it, even with just a small amount. If there is an interruption in Truck transportation, manufacturing industries shake too, because among other effects probably, the spare parts don't arrive in time. Therefore, if we do not consider these transactions, we lose some essential propagation mechanisms.

These small monetary transactions not only directly contribute to the propagation, as Truck transportation directly supports other industries, but they also function as intermediaries. If we cut them, we ignore all the propagation mechanisms along the chain. The Truck transportation disturbance might impact the manufacturing industries, but this shock might propagate even further in smaller proportion to the third-forth industries connected to the manufacturing ones.

Another aspect that is quite important but we didn't capture through this analysis is that these interdependencies are in no way linear. For example, the Truck transportation industry might contribute only 5\% to Car manufacturing. Still, if the Truck transportation industry stops, the other sector falls to zero without access to that crucial part. This limitation of our study can be overcome by examining exactly this resilience aspect in future work.

This study has shown that the threshold value $\zeta$ chosen makes a significant difference in the production network's topology; therefore, we must carefully consider and keep in mind the limitations of the research made on production networks using a particular threshold $\zeta$.

As a further step of the research, it would be reasonable to explore the propagation mechanisms of particular industries and analyse to what extent the network properties explain the growth of different industries.

\begin{backmatter}

\section*{Availability of data and materials}
  The datasets generated and analysed during the current study are available here:  https://github.com/molnareszter/Topology-dependence-of-propagation-mechanisms-in-production-network

\section*{Competing interests}
  The authors declare that they have no competing interests.

\section*{Author's contributions}
EM and DC designed the research. EM collected, preprocessed, and visualized the data. EM and DC analyzed the data and interpreted the results. EM wrote the manuscript. DC reviewed and revised the manuscript.

\section*{Acknowledgements}
  We would like to thank Levente Sz\'{a}sz for his valuable advice.
  
\section*{Funding}
  EM acknowledges the support of the Leverhulme Doctoral Scholarships Programme in Material Social Futures (grant number DS-2017-036).
 
\bibliographystyle{bmc-mathphys} 
\bibliography{article}

\end{backmatter}

\end{document}